\documentclass[a4paper,11pt]{article}
\pdfoutput=1 

\usepackage{jheppub} 

\usepackage[T1]{fontenc} 
\usepackage[normalem]{ulem}
\usepackage{indentfirst}
\usepackage{subfig}
\usepackage{xcolor}

\title{Dark Matter Freeze-In during Warm Inflation and the Seesaw Mechanism}

\author[a]{Rayff de Souza,}\emailAdd{rayffsouza@on.br}
\author[a]{Jamerson G. Rodrigues,}\emailAdd{jamersoncg@gmail.com}
\author[b,a]{Clarissa Siqueira,}\emailAdd{siqueira.cms@gmail.com}
\author[a]{Felipe B. M. dos Santos,}\emailAdd{fbmsantos@on.br}
\author[a]{Jailson Alcaniz}\emailAdd{alcaniz@on.br}
\affiliation[a]{Observatório Nacional, 20921-400, Rio de Janeiro - RJ, Brazil}
\affiliation[b]{Universidade Federal de Campina Grande, Unidade Acad\^emica de Física, Campina Grande - PB, Brazil}

\abstract{A compelling way to address the inflationary period is via the warm inflation scenario, where the interaction of the inflaton field with other degrees of freedom affects its dynamics in such a way that slow-roll inflation is maintained by dissipative effects in a thermal bath. In this context, if a dark matter particle is coupled to the bath due to non-renormalizable interactions, the observed dark matter abundance may be produced during warm inflation via ultra-violet freeze-in. In this work, we propose applying this scenario in the framework of a $U(1)_{B-L}$ gauge extension of the Standard Model of Particle Physics, where we also employ the seesaw mechanism for generating neutrino masses.}

\begin{document} 
\maketitle
\flushbottom

\section{Introduction}\label{sec1}
One of the most well-established components of the universe's energy budget is dark matter (DM), whose evidence for its existence is provided by several observational results that include galaxy rotation curves, dynamics of galaxy clusters, gravitational lensing, and the cosmic microwave background (CMB) \cite{Bertone:2004pz}. Despite the firm confidence in its existence, the DM nature and production mechanism remain unknown, regardless of significant efforts to detect its possible non-gravitational interactions \cite{LZ:2022lsv,XENON:2023cxc}. A general method to produce DM in the early universe is to consider an interaction between the DM particle with the primordial thermal bath, where the behavior and rate of DM production are dictated by the strength of its coupling \cite{Gondolo:1990dk,Elahi:2014fsa}. If the interaction is weak enough, DM may never reach thermal equilibrium, allowing its number density to gradually increase over time, in a process known as DM freeze-in \cite{Hall:2009bx}. This DM-bath coupling may be due to renormalizable interactions with small coupling constants, such as infra-red (IR) freeze-in, where the DM abundance depends on low temperatures near the DM mass scale \cite{Hall:2009bx}. On the other hand, the coupling may be provided by non-renormalizable interactions with a heavy mass scale, as in the ultra-violet (UV) freeze-in case, with the DM abundance sensitive to the highest bath temperatures reached during cosmic evolution \cite{Elahi:2014fsa}. 

In addition, the concept of cosmological inflation \cite{Guth:1980zm,Starobinsky:1980te,Linde:1981mu}, where the universe undergoes an early stage of accelerated expansion before the hot big bang evolution, has also received further support from the latest CMB observations made by the Planck Collaboration \cite{Planck:2018vyg,Planck:2018jri}. More specifically, tighter constraints have been imposed on the primordial power spectra of cosmological perturbations, which carry the bulk of information from the inflationary epoch. The scalar spectrum is expected to be nearly scale-invariant at more than $2\sigma$, with statistics well described by a Gaussian distribution. These findings are consistent with the simplest theoretical constructions of inflation, where a scalar field $\phi$, dubbed the inflaton, has negligible interactions and slowly rolls down its potential $V(\phi)$, driving cosmic acceleration via slow-roll inflation \cite{Guth:1980zm,Starobinsky:1980te,Linde:1981mu}. However, a concrete embedding of inflation in realistic theoretical frameworks has proven to be challenging, due to stronger limits placed on the available parameter space of many inflationary models \cite{Martin:2024qnn}. In addition, stringent constraints on the amplitude of primordial tensor modes, via observations of the B-mode spectra of CMB polarization \cite{BICEP:2021xfz,Tristram:2021tvh}, have ruled out the simplest inflationary potentials, such as chaotic or monomial inflation \cite{Planck:2018jri}.

In light of the shortcomings of well-motivated inflationary models, an intriguing hypothesis is to consider that the interactions of the inflaton with thermal fields might play a role in the dynamics of inflation. In this instance, dissipation of energy from the inflaton to radiation may sustain a heat bath throughout the inflationary period, while also providing the friction to maintain the inflaton in the slow-roll regime. After the inflaton had dissipated its energy away, inflation would end in a radiation-dominated universe, with little to no need for a separate reheating epoch. Such a scenario is called Warm Inflation (WI) \cite{Berera:1995ie,Berera:2008ar} and its phenomenology has been extensively explored as a viable extension of the canonical inflationary picture \cite{Bastero-Gil:2009sdq,Arya:2017zlb,Bastero-Gil:2017wwl,Arya:2018sgw,Berghaus:2019whh,Das:2020xmh,Santos:2022exm,Montefalcone:2022jfw,Berera:2023liv}. Moreover, models of WI generally produce very low values for the tensor-to-scalar ratio $r$ \cite{Benetti:2017juy}, which helps to restore concordance of monomial inflation with current data \cite{Santos:2024pix}. 

Recently, the authors in \cite{Freese:2024ogj} have developed a novel interplay between DM production and inflationary cosmology. They showed that, if a DM-bath coupling of the UV freeze-in type is active during a warm inflationary period, a large abundance of DM is generally produced, with the resulting DM yield typically larger than the standard radiation-dominated UV freeze-in, for the same reheating temperature. They also demonstrated that this warm inflation UV freeze-in (WIFI) is robust for a variety of mass dimensions of the operator that effectively couples DM to the thermal bath.

Since the WIFI framework involves the addition of a DM candidate to the Standard Model of Particle Physics (SM), it is then natural to investigate whether such a mechanism is also able to explain the generation of neutrino masses, which represents one of the most promising signatures of beyond the standard model physics. In this sense, we attempt to embed the WIFI mechanism into a specific extension of the SM, where neutrino masses are generated via the seesaw mechanism. More specifically, we employ a $U(1)_{B-L}$ gauge extension of the SM~\cite{Buchmuller:1991ce,King:2004cx,Basso:2008iv}. We explore the possibility of a fermionic DM component, whose production mechanism realized during inflation is mediated by the new gauge boson, $Z^\prime$. The hierarchies of mass scales imposed by both the WIFI and seesaw mechanisms are explained by the spontaneous breaking of the $B-L$ gauge symmetry. Furthermore, the constraints imposed by the Bayesian analysis of warm inflation scenarios carried out in \cite{Santos:2024pix} are considered and the phenomenology of the proposed model is investigated.

This work is organized as follows. In Section \ref{sec2}, we outline the main aspects of the dynamics of warm inflation, as well as a general description of the WIFI mechanism. In Section \ref{sec3}, we present a particle physics construction of WIFI and also a description of the seesaw mechanism. We present our results in Section \ref{sec4}, for the WIFI implications on our model's scalar and Yukawa sectors, and DM production. We end this paper by presenting our concluding remarks in Section \ref{sec5}.

\section{Elements of Warm Inflation}\label{sec2}
As opposed to the standard cold inflationary picture, WI considers a coupling of the inflaton with thermal fields, such that radiation is continuously produced through dissipation during inflationary expansion. This interaction is usually represented by a dissipation coefficient $\Upsilon$, which depends on the specific interactions of the inflaton with the environmental fields in the bath \cite{Bastero-Gil:2010dgy}. With the inclusion of the dissipation term, the WI background equations read:
\begin{equation}\label{background eq 1}
    \ddot\phi + 3H\dot\phi + V_{,\phi} = -\Upsilon\dot\phi\; ,
\end{equation}
\begin{equation}
    \dot\rho_r + 4H\rho_r = \Upsilon\dot\phi^2\; ,\label{background eq 2}
\end{equation}
\begin{equation}\label{background eq 3}
    H^2 = \frac{\rho_\phi + \rho_r}{3M_p^2}\; , 
\end{equation}
where $\rho_\phi = \dot\phi^2/2 + V(\phi)$ and $\rho_r = (\pi^2 g_\star(T)/30)T^4$ are the energy densities in the inflaton field and radiation, respectively, $g_\star(T)$ is the effective number of relativistic degrees of freedom and $M_p = 1/\sqrt{8\pi G} \approx 2.4 \times 10^{18}$ GeV is the reduced Planck mass.

The effectiveness of the dissipation of energy from the inflaton to the thermal bath is parameterized by the dimensionless dissipation parameter:
\begin{equation}
    Q \equiv \frac{\Upsilon}{3 H}\; .
\end{equation}

In the weak dissipative regime, $Q \ll 1$, so that the Hubble friction still dominates the damping of the inflaton's motion. On the other hand, in the strong dissipative regime, $Q \gg 1$, dissipation is the strongest effect in keeping the inflaton in slow-roll. Generally, WI can happen in one of the dissipation regimes or even transition from one to the other, depending on the dissipative coefficient and inflationary potential. Moreover, we assume that thermalization of the radiation bath is rapidly achieved, resulting in an almost constant bath temperature that remains larger than the Hubble expansion\footnote{This is also a requirement for safely computing the dissipative coefficient in the flat spacetime approximation \cite{Bastero-Gil:2010dgy}.}. In addition, due to the coupling between the inflaton and radiation at the perturbative level, the primordial scalar power spectrum is modified with  temperature-dependent terms \cite{Hall:2003zp,Graham:2009bf,Bastero-Gil:2011rva}.

As a representative example of the WI scenario, we can consider a simple quartic inflationary potential, $V(\phi) = \lambda\phi^4/4$. Such scalar potential is ruled out in the canonical inflationary picture \cite{Planck:2018jri} but can be brought back into concordance in the WI framework \cite{Santos:2024pix,Bastero-Gil:2017wwl,Arya:2017zlb,Benetti:2016jhf,Kumar:2024hju}. For the dissipative coefficient, we consider both a linear ($\Upsilon \propto T$) and a cubic ($\Upsilon\propto T^3/\phi^2$) dependence on the bath temperature. The linear case is obtained in the Warm Little Inflaton model \cite{Bastero-Gil:2016qru}, in which WI is successfully realized with a small number of fields, while the cubic coefficient was one of the first proposals of WI, motivated by supersymmetry arguments \cite{Bastero-Gil:2012akf}. For this combination of potential and dissipative coefficients, recent constraints on the primordial spectrum of WI with CMB data seem to favor the weak dissipative regime, with the best-fit for the dissipation parameter at the Planck reference scale horizon crossing being $\log Q_\star = -2.11$ for the linear coefficient and $\log Q_\star = -2.32$ for the cubic case \cite{Santos:2024pix}. Using $\log Q_\star = -2$, which is inside the $1\sigma$ limit for both coefficients \cite{Santos:2024pix}, and imposing the normalization of the spectrum at the reference scale to be $\Delta_\mathcal{R}^2 \sim 2.1\times 10^{-9}$, we fix the potential quartic coupling in both cases to $\lambda_\text{linear}\sim 3.3\times 10^{-15}$ and $\lambda_\text{cubic} \sim 2.7\times 10^{-14}$. Then, we can iteratively solve the background Eqs.~\eqref{background eq 1} - \eqref{background eq 3} in terms of the number of e-folds $d N_e = H d t$, until the condition for the end of inflation $\epsilon_H \equiv - \frac{\dot H}{H^2} = 1$ is achieved after $\sim 50 - 60$ e-folds, which signals the end of inflation after relevant CMB scales crossed the horizon.

\begin{figure*}[t]
\centering
\includegraphics[scale=0.7]{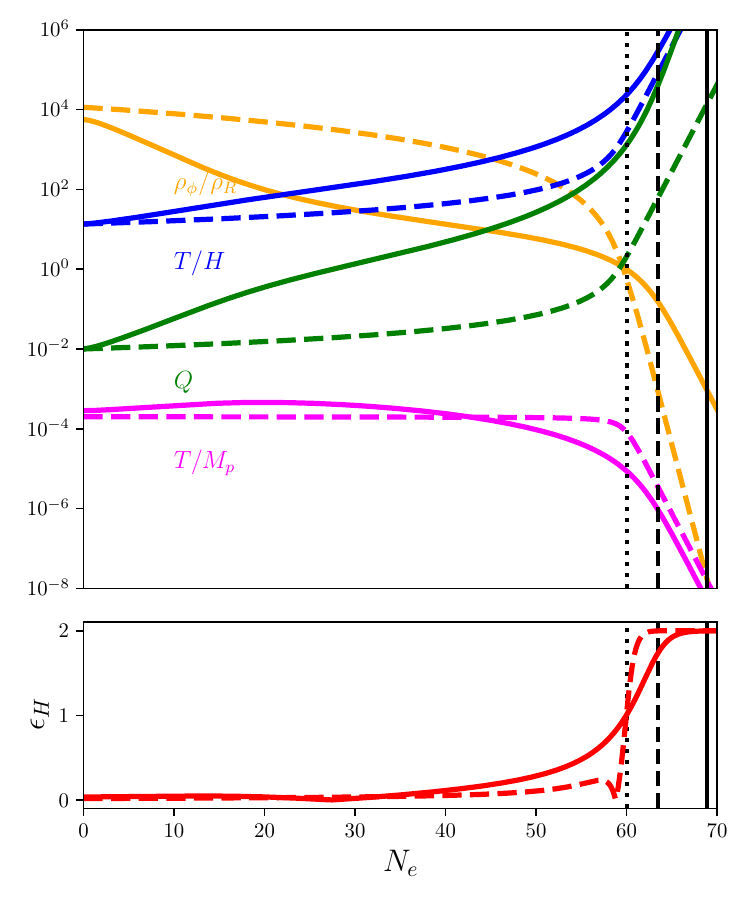}
\caption{Background evolution of a few WI dynamical quantities with a quartic scalar potential, with CMB scales crossing the horizon 60 e-folds before the end of inflation ($\epsilon_H = 1$, dotted vertical line). The cases for a cubic (linear) dissipation coefficient are plotted in solid (dashed) curves. The solid and dashed vertical lines indicate the time of reheating ($\epsilon_H = 2$) for the cubic and linear models, respectively.}
\label{fig:1}
\end{figure*}

Figure~\ref{fig:1} shows the evolution of relevant dynamical quantities for both dissipation coefficients in the case of $N_{e,\star} = 60$. As one can see, the bath temperature does not vary much throughout inflation and stays larger than the Hubble scale, at an average of $\bar T_\text{linear} \sim 4.6\times 10^{14}$ GeV and $\bar T_\text{cubic} \sim 6.8\times 10^{14}$ GeV. Also, WI happens in the weak dissipative regime at the horizon crossing of CMB scales but ends with $Q \sim 1$ for the linear coefficient and deep in the strong dissipative regime for the cubic case. As expected, the energy density in the inflaton field dominates the inflationary expansion, but quickly drops below the radiation energy density towards the end of inflation. In the lower panel of Figure~\ref{fig:1}, we show the evolution of the Hubble slow-roll parameter $\epsilon_H$. The reheating point sets the onset of the radiation-dominated (RD) regime, which is taken as the limit of $\epsilon_H$ when $\rho_r \gg \rho_\phi$, yielding $\epsilon_H|_\text{reh} = 2$. For the linear dissipative coefficient, RD starts at $N_{e\text{,reh}} \sim 63$, while $N_{e\text{,reh}} \sim 69$ for the cubic coefficient.

\subsection{Warm Inflation Freeze-In}\label{subsec2.1}
During DM freeze-in, the DM abundance is slowly brought up to the present-day value as a result of the coupling between DM and the thermal bath in the early universe. In standard inflationary cosmology, inflation happens in a supercooled low-entropy state, such that any thermal production of DM has to be assigned to the post-inflationary expansion. Since in WI a persistent thermal bath is maintained all throughout, one can then ponder the possibility that DM may have been thermally produced during inflation. This hypothesis was investigated in \cite{Freese:2024ogj} in the context of UV freeze-in, where the DM-thermal bath coupling is due to non-renormalizable operators with a heavy mass scale. In this case, the Boltzmann equation for the DM number density $n_\chi$ reads \cite{Elahi:2014fsa}
\begin{equation}\label{boltzmann}
    \dot n_\chi + 3 H n_\chi = \frac{T^{2 n + 4}}{\Lambda^{2 n}}\; .
\end{equation}

The right-hand side represents an effective interaction between DM and the radiation bath, coming from an operator with mass dimension $n + 4$, with the scale $\Lambda$ being the cutoff of the effective description, which remains valid as long as $T < \Lambda$.

In terms of the DM yield $Y_\chi \equiv n_\chi/s$, where $s = (2\pi^2/45)g_\star(T) T^3$ is the entropy density of the thermal bath, Eq.~\eqref{boltzmann} can be written as\footnote{In the following, we assume that $d g_\star(T)/d T = 0$ so that $g_\star(T) = g_\star$.}
\begin{equation}\label{boltzmann 2}
    \dot Y_\chi + 3\left(\frac{\dot T}{T} + H\right)Y_\chi = \left(\frac{45}{2\pi^2 g_\star}\right)\frac{T^{2 n + 4}}{\Lambda^{2 n}}\; .
\end{equation}

In the standard radiation-dominated freeze-in, the adiabatic condition, which states that entropy is conserved and there is no decay of particles into radiation, implies that $T \propto 1/a$ so that $\dot T/T = -H$ and the parenthesis in the left-hand side of Eq.~\eqref{boltzmann 2} vanishes \cite{Kolb:1990vq}. However, in WI, the adiabatic condition does not hold since radiation is continuously produced, so the temperature stays approximately constant - see Figure~\ref{fig:1}. This marks an important distinction between WIFI and the radiation-dominated freeze-in. Thus, in terms of the number of e-folds $N_e$, the Boltzmann equation reads
\begin{equation}\label{boltzmann final}
    Y'_\chi + 3\left( \frac{T'}{T} + 1\right)Y_\chi = \left(\frac{45}{2\pi^2 g_\star}\right)\frac{T^{2 n + 4}}{\Lambda^{2 n} H}\; ,
\end{equation}
where $Y'_\chi=dY_\chi/dN_e$.
Therefore, the DM yield is sourced by a function of the bath temperature and the Hubble scale, depending on the effective operator exponent $n$. We can then plug Eq.~\eqref{boltzmann final} in the differential system given by Eqs.~\eqref{background eq 1} - \eqref{background eq 3}, so that the DM abundance is produced according to the WI background evolution. After inflation, the DM final yield will depend on the heavy mass scale $\Lambda$. Hence, we can fix $\Lambda$ by matching the equilibrium value of $Y_\chi$ given by the coupled background system to the yield needed to account for the observed DM abundance $\Omega_\chi h^2 \simeq 0.1200\pm 0.0012$ \cite{Planck:2018vyg}. For a given DM mass, $Y_{\chi,\text{eq}} \sim 6.45\times 10^{-10}\left( \frac{1\text{ GeV}}{m_\chi} \right)$, using the most recent data from PDG \cite{ParticleDataGroup:2024cfk}.

In reference \cite{Freese:2024ogj}, the authors performed an extended analysis of the DM yield evolution in WIFI for the effective operator mass dimension up to $n = 5$, as well as other values for the dissipation parameter at horizon crossing $Q_\star$. They also showed that the WIFI mechanism always produces a greater DM yield than the standard radiation-dominated UV freeze-in for the same reheating temperature, defined as $T_\text{reh} = T|_{\epsilon_H = 2}$, where the ratio between the two increases exponentially with $n$. 

\section{Particle Physics Setup}\label{sec3}

In this section, we aim to construct a particle physics realization of the WIFI mechanism. With this purpose and according to \cite{Freese:2024ogj}, two conditions must be addressed. The first one secludes the inflaton field and the DM candidate to distinct sectors, forbidding their direct coupling. The second imposes a hierarchy between the mass scale of DM and the energy scale of the UV physics associated to the production mechanism, i.e., $m_{DM} < T < \Lambda$, where $T$ is the temperature of the radiation bath. Inside these limits, the direct decay of the inflaton into DM is avoided, as well as the back-reaction of DM into particles of the thermal bath.

In particular, we explore the possibility of a fermionic DM candidate, $\chi$, once a scalar field would inherit a quartic coupling to the inflaton, potentially breaking the first conjecture of WIFI. The hierarchy imposed from the second condition is obtained following the seesaw mechanism for generating neutrino masses. In our scenario, the DM production throughout inflation is mediated by a new gauge boson, $Z^\prime$. All these pieces will be arranged in the framework of the $U(1)_{B-L}$ gauge extension of the SM. In this set up, the origin of the energy scale $\Lambda$, the smallness of neutrino masses, and the hierarchy of the WIFI mechanism are explained following the spontaneous breaking of the $B-L$ symmetry.

\subsection{Seesaw Mechanism}\label{subsec3.1}

The general sobriquet of seesaw mechanisms aggregates a set of mathematical artifices in order to provide an elegant description for the smallness of the neutrinos masses~\cite{Minkowski:1977sc,Yanagida:1979as,GellMann:1980vs,Mohapatra:1979ia,Schechter:1980gr}.  The inverse seesaw mechanism (ISS) \cite{Mohapatra:1986aw,Mohapatra:1986bd}, although not the most economical of its realizations but certainly one of the most phenomenologically interesting, requires the addition of three right-handed neutrinos $N_i$, and three sterile fermions $\chi_i$ to the SM content. New symmetries are evoked in order to allow the following bilinear terms,
\begin{equation}
    \mathcal{L} \supset -  \bar{\nu}_L m_D N -  \bar{\chi}^c M N - \frac{1}{2} \bar{\chi}^c \mu \chi + h.c.,
\end{equation}
where $\nu_L$ are the active neutrinos of the SM and $m_D$, $M$ and $\mu$ are generic $3\times 3$ mass matrices.  These terms can be arranged in a $9\times 9$ matrix, which under the basis $(\nu_L,\, N^c, \, \chi^c)$ assumes the form,
\begin{eqnarray}
    M_\nu = \left(\begin{array}{ccc}
0 & m^T_D & 0 \\ 
m_D & 0 & M^T \\
0 & M & \mu\end{array}\right).
\end{eqnarray}
The corresponding fields written in terms of their mass basis are obtained through the diagonalization by blocks procedure \cite{Schechter:1981cv,Hettmansperger:2011bt,Dias:2012xp}.   Considering the hierarchy $\mu \ll m_D \ll M$, one may obtain the three standard neutrinos with mass matrix of the form,
\begin{equation}
    m_\nu = m^T_D(M^T)^{-1}\mu M^{-1} m_D,
\end{equation}
doubly suppressed by the factor $\text{Det}[(M^T)^{-1}\mu M^{-1}]$, which warrant sub-eV mass to neutrinos for reasonable values of the inputs, e.g. $\mu \sim$ keV, $m_D \sim 100$ GeV and $M \sim$ TeV. Meanwhile, the six sterile fermions acquire masses in the same scale and proportional to $M$. 

In order to reproduce the new hierarchy imposed by the WIFI mechanism, $m_{DM} < T < \Lambda$,  while preserving the correct magnitude of masses of the standard neutrinos,  we propose a new bilinear term to the Lagrangian in order to break the degeneracy of the sterile sector,
\begin{equation}
    \mathcal{L} \supset -  \bar{\nu}_L m_D N  -  \bar{N}^c \mu_N N -  \bar{\chi}^c M N - \frac{1}{2} \bar{\chi}^c \mu \chi + h.c., \label{Eq:Yu_nu}
\end{equation}
where the corresponding $9\times 9$ mass matrix now reads,
\begin{eqnarray}
    M^\prime_\nu = \left(\begin{array}{ccc}
0 & m^T_D & 0 \\ 
m_D & \mu_N & M^T \\
0 & M & \mu\end{array}\right).
\end{eqnarray}
This new mass scale is associated with the energy scale of the mediator responsible for the freeze-in production of DM, $\mu_N \propto \Lambda$.

\subsection{The Model} \label{Model}

In this work, we explore a realization of the ISS in the $U(1)_{B-L}$ extension of the SM \cite{Buchmuller:1991ce,King:2004cx,Basso:2008iv}\footnote{See also \cite{Khalil:2010iu} for a different realization of the ISS in the framework of the $B-L$ extension.}.
In particular, we extend the SM content of fields with two new complex scalars, $\phi$ and $\sigma$, and six new sterile fermions, $N^i$ and $\chi^i$, with $i=1,2,3$\footnote{In order to cancel the gravitational and gauge anomalies coming from the new abelian symmetry one should also include three additional sterile fermions with $B-L$ charge $-1$. We ignore the role played by these new fields by imposing a parity symmetry on the Lagrangian with all other fields even.}, where the lightest $\chi$ fermion will be our DM candidate. All of them singlet under the canonical SM symmetries and with $B-L$ charges assigned following the Yukawa terms,
\begin{equation}
    \mathcal{L} \supset - Y^{ij}_\nu \bar{L}_i \tilde{H} N_j - Y^{ij}_N \bar{N}^c_{i}\phi N_j - Y^{ij}_M \bar{\chi}^c \sigma N - \frac{1}{M^3_\mu} \bar{\chi}^C_i \sigma^4 \chi_j, \label{Eq:Yu_nu2}
\end{equation}
corresponding to $Y^{\sigma,\phi,\chi,N}_{\text{\tiny{B-L}}} = -1, +2, +2, -1$.  Also, $L=(\nu \,\,,\,\, e)_L^T$ and $\tilde{H}=(H^0 \,\,,\,\, -H^-)^T$ are the standard leptons and the conjugated Higgs doublet, respectively. In light of the requirements of the WIFI mechanism, namely to avoid the coupling between the inflaton and DM candidate, we choose the $\phi$ scalar to play the role of the inflaton field.

Once the neutral scalar fields settle in their vacuum expectation values (vev), the $B-L$ abelian symmetry, as well as the electroweak symmetry, become hidden by the vacuum structure 
\begin{equation}
    H^0,\, \phi,\, \sigma \rightarrow \frac{1}{\sqrt{2}}(v_{h,\, \phi,\, \sigma} + R_{h,\, \phi,\, \sigma} + iI_{h,\, \phi,\, \sigma}),\label{Eq:vev}
\end{equation}
and the majority of the fields, except the photon, acquire mass terms. In particular, the $U(1)_{B-L}$ symmetry breaking process is triggered by the vev of the $\phi$ field at a high energy scale, $v_\phi \gg 100$ GeV. The gauge boson $Z^\prime$, associated with the $B-L$ symmetry, becomes massive through the usual Higgs mechanism, with the kinetic Lagrangian allowing the mass,
\begin{equation}
    m_{Z^\prime} = g_{\text{\tiny{B-L}}}v_\phi,
\end{equation}
where $g_{\text{\tiny{B-L}}}$ is the gauge coupling. The remnant global symmetry of $(-1)^L$ is preserved in the system throughout the $B-L$ breaking process, with $L$ being the lepton number. This symmetry is broken at a much lower energy scale by the vev of $\sigma$, $100\, \text{GeV}\, < v_\sigma \ll v_\phi$.

Finally, the SM electroweak symmetry is broken by the Higgs vev, $v_h \sim 10^2$ GeV, and the Yukawa sector of the neutrinos enables the terms in Eq.~\eqref{Eq:Yu_nu},
\begin{equation}
     \mathcal{L} \supset -  \bar{\nu}_L m_D N  -  \bar{N}^c \mu_N N -  \bar{\chi}^c M N - \frac{1}{2} \bar{\chi}^c \mu \chi + h.c.,
\end{equation}
where $m_D = Y_\nu v_h/\sqrt{2}$, $\mu_N = Y_N v_\phi/\sqrt{2}$, $M = Y_M v_\sigma/\sqrt{2}$ and  $\mu = v^4_\sigma/M^3_\mu$. The non-renormalizable operator in Eq.~\eqref{Eq:Yu_nu2} is generated radiatively \cite{Khalil:2010iu,Ma:2009gu}, corroborating to the hierarchy enforced by the conjunction of ISS and WIFI mechanisms, $\mu < M \ll \mu_N$.

The scalar potential enabling such a vacuum configuration assumes the form,
\begin{eqnarray}
    V(H, \phi, \sigma) = && m^2_h H^\dagger H +  m^2_\phi \phi^\dagger \phi + m^2_\sigma \sigma^\dagger \sigma  + \lambda_h (H^\dagger H)^2 + \lambda_\phi (\phi^\dagger \phi)^2 \nonumber \\
    && + \lambda_\sigma (\sigma^\dagger \sigma)^2 + \lambda_1 (H^\dagger H)(\sigma^\dagger \sigma) +  \lambda_2 (H^\dagger H)(\phi^\dagger \phi) + \lambda_3 (\phi^\dagger \phi)(\sigma^\dagger \sigma) \nonumber \\
    && + m_{nh}\phi \sigma^2 + h.c., \label{Eq:Pot_1}
\end{eqnarray}
following the underlying gauge symmetries. The existence of a minimum in the potential requires that its first derivatives, relative to the neutral scalar fields, to vanish, resulting in three constraint equations,
\begin{eqnarray}
    &&  2m^2_\sigma + 2\lambda_\sigma v^2_\sigma + \lambda_3 v^2_\phi   + \lambda_1 v^2_h +2\sqrt{2} m_{nh} v_\phi  =  0,\nonumber \\
    && 2m^2_h + 2 \lambda_h v^2_h + \lambda_1 v^2_\sigma + \lambda_2 v^2_\phi = 0  , \label{Eq:Min} \\
    && 2m^2_\phi v_\phi + 2 \lambda_\phi v^3_\phi + \lambda_3 v_\phi v^2_\sigma + \lambda_2 v_\phi v^2_h + \frac{m_{nh}}{\sqrt{2}}v_\sigma^2 =  0. \nonumber
\end{eqnarray}
Given the difference in the magnitude of the vevs, the terms with higher powers of $v_\phi$ tend to dominate the sums above, resulting in typically large and negative values for $m^2_\phi$, $m^2_h$ and $m^2_\sigma$, as usual for a Spontaneous Symmetry Breaking process (SSB).

\begin{figure}
    \centering
    \includegraphics[width=0.35\linewidth]{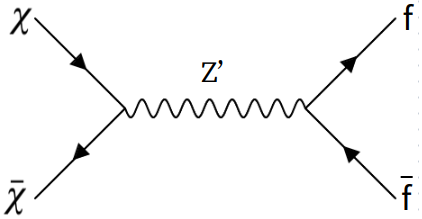}
    \caption{The Feynman diagram that drives the production of the DM particle through the freeze-in mechanism during warm inflation. $f$ means all fermions in the SM, including quarks and leptons.}
    \label{sec3:diag}
\end{figure}

In this framework, we propose that the lightest $\chi$ fermion constitutes the DM candidate, which provides the DM relic density through the freeze-in mechanism. The Lagrangian, which couples the DM field $\chi$ to the new gauge boson $Z^{\prime}$, is given by
\begin{equation}
    \mathcal{L}_{\chi} = i \bar{\chi} \left( \partial_\mu - i g_{\text{\tiny{B-L}}} Y^{\chi}_{\text{\tiny{B-L}}} Z^{\prime}_\mu \right)\gamma^\mu \chi,
\end{equation}
which leads to the dominant contribution to the DM relic density. The SM fermions follow the same gauge structure, leading to the two-to-two fermionic interaction involving the DM fermion and the SM ones. The Feynman diagram is provided in Figure~\ref{sec3:diag}.

\subsection{Corrections to the Inflaton Potential}\label{Corrections}

The high temperatures of the thermal bath during the warm inflationary period may lead to sizable thermal corrections to the inflaton potential, threatening the flatness necessary for slow-roll inflation. These thermal corrections can be quantified by the Helmholtz free energy density, $F(h,\phi,\sigma,T)=V(h,\phi,\sigma,T)$, where the relevant contribution may be written as \cite{Sher:1988mj,Cline:1996mga}
\begin{equation}
    \Delta V \supset \frac{m^2(\alpha)}{24}T^2 + \frac{m^4(\alpha)}{16\pi^2}\ln{\left(\frac{\mu^2}{T^2}\right)}, \label{Eq:deviation}
\end{equation}
for $m \ll T$, where $m(\alpha)$ is the effective quadratic mass for scalars $\alpha = H,\phi, \sigma$ and $\mu$ is the renormalization scale.

From the Yukawa sector given by Eq.~\eqref{Eq:Yu_nu2} and the discussion of the SSB process carried in the Sec.~\ref{Model}, we note that the thermal corrections arising from the interaction between the right-handed neutrinos ($N$) and the inflation field ($\phi$) are suppressed by a Boltzmann exponential factor, given the large mass associated with these particles, typically $m_N > T$ \cite{Bastero-Gil:2010dgy}. In this sense, the high energy scales of the seesaw mechanism act to armor the inflaton potential against fermionic thermal corrections. With this same reasoning, the condition that $m_{Z^{'}}  > T$, which allows the diagram in Figure \ref{sec3:diag} to be described by an effective DM-bath interaction, also protects the inflaton potential from harmful bosonic corrections. Further, thermal corrections coming from the scalar interactions of the inflaton in the scalar potential, given by Eq.~\eqref{Eq:Pot_1}, are highly suppressed by the small value of the mixing couplings, where $\lambda_2 = \lambda_3 \sim 10^{-8}$, following the discussion below the Eqs.~ \eqref{Eq:Min}.

From the perspective of the SSB processes, the thermal corrections also displace the minimum of the scalar potential, potentially recovering some of the symmetries discussed above through the inflationary period. The representative contribution for the minimum conditions in Eq.~\eqref{Eq:Min} amounts to
\begin{equation}
    \langle \Delta V \rangle_0 \propto \frac{1}{24} (\lambda v)_{h,\phi,\sigma}\, T^2  . 
\end{equation}
For the typical values of $v_{h,\phi,\sigma}$ and $T$ introduced previously, one may find that such terms are at least three orders of magnitude lower than the dominant contribution, $ 2 \lambda_\phi v^3_\phi$. The main conclusion is that the $B-L$ symmetry is broken in the thermal bath throughout the warm inflation process, while the discrete $(-1)^L$ and the electroweak symmetry are preserved. Consequently, the SSB scale $\Lambda$ is the relevant one in the freeze-in process.

Moreover, we consider that the Coleman-Weinberg term in Eq.~\eqref{Eq:deviation}, coming from radiative corrections to the potential, is also sub-leading in the inflationary regime. This amounts to consider that the Renormalization Group $\beta$-function of the inflaton quartic coupling is small during inflation, $\beta_\lambda \ll 1$. This may be produced by a canceling between bosonic and fermionic contributions.

Finally, we stress that we stay agnostic about the inflaton interactions responsible for sustaining the thermal bath through dissipation, delegating these to the underlying microphysical model of WI \cite{Bastero-Gil:2016qru,Bastero-Gil:2012akf,Berera:2023liv}. However, we have checked that the inflaton couplings in Sec.~\ref{Model} do not result in harmful corrections to the inflaton potential, which implies that we can safely consider the tree-level potential $V(\phi) \sim \lambda_\phi \phi^4$ to be the relevant term during WI dynamics.

\section{Results}\label{sec4}

In this Section, we analyze the ultraviolet freeze-in DM production during warm inflation for the model presented in Sec. \ref{Model}. Namely, we consider a DM particle effectively coupled to the thermal bath through a mass dimension 6 non-renormalizable operator ($n = 2$). For the WI scenario, we consider both the linear and cubic models of Sec. \ref{sec2} in the weak dissipative regime, with $\log Q_\star = -2$.

\begin{figure*}[t]
\centering
\includegraphics[width=1\columnwidth]{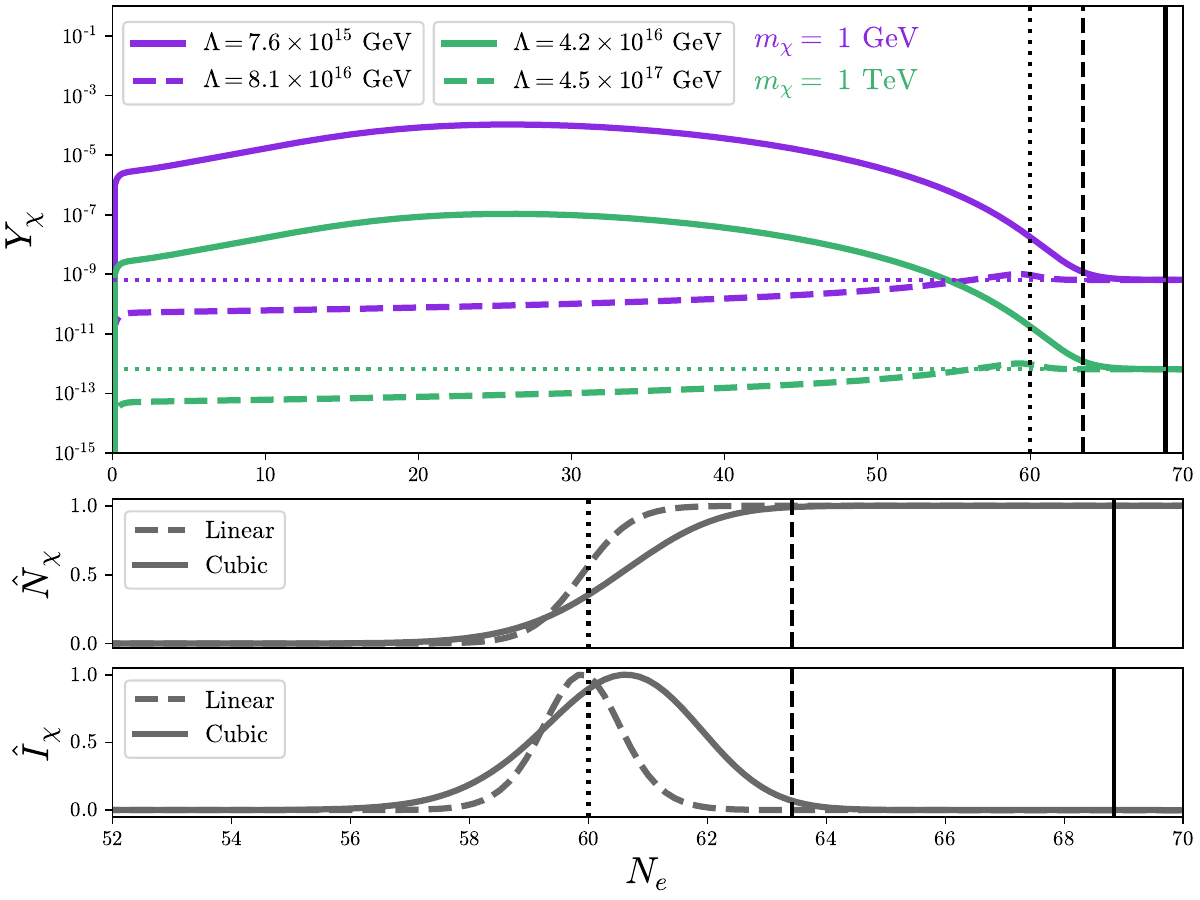}
\caption{Upper panel: Evolution of the DM yield for $m_\chi = $ 1 GeV (purple) and $m_\chi = $ 1 TeV (green), for both cases of the temperature dependence of the WI dissipation coefficient, linear (dashed) and cubic (solid). The values of the cutoff scale $\Lambda$ for each curve are displayed on the legend, with the horizontal dotted lines denoting the DM equilibrium yield for each DM particle mass.
Middle and lower panels: Evolution of the normalized DM comoving number density $\hat{N}_\chi$ and its normalized derivative $\hat{I}_\chi$, for both the linear and cubic models. In all panels, the dotted, dashed and solid vertical lines correspond to the end of inflation and the reheating moment for the linear and cubic cases, respectively.}
\label{fig:3}
\end{figure*}

In the upper panel of Figure \ref{fig:3}, we show the evolution of the DM yield for the representative cases of 1 GeV and 1 TeV masses of the DM particle. The $\Lambda$ scaled is matched so that the final yield results in the observed DM abundance. In the middle and lower panels, we plot the evolution of the comoving DM number density $N_\chi \equiv n_\chi e^{3N}$ as well as its derivative $I_\chi = d N_\chi/ d N_e$, both normalized so their maximum values are equal to 1. Since $\Lambda$ is always a high energy scale and does not change much for the DM masses considered, the comoving DM number density will only be affected by $n$ (which, in our model, is fixed to $n = 2$) and the evolution of the WI bath temperature $T$ and Hubble scale $H$, as can be seen from Eq. \eqref{boltzmann}. Therefore, the curves for $\hat{N}_\chi$ and $\hat{I}_\chi$ for the 1 GeV and 1 TeV DM masses are superimposed on top of each other, depending only on the choice of the WI dissipation coefficient. 

Also, the middle and lower panels allow us to better investigate the production of DM in the periods before and after the end of inflationary expansion. In both the linear and cubic WI models, the comoving DM number density starts to increase in the few e-folds before the end of inflation. For the linear model, DM production peaks shortly before the end of inflation and roughly 1 e-fold afterwards for the cubic coefficient. In both cases, the relic value of  $N_\chi$ is fully established before the onset of RD, similar to the results of \cite{Freese:2024ogj} for higher values of $n$.

In order to obtain the observed DM abundance, we have to fix the cutoff scale $\Lambda$ to the values displayed on top of the upper panel of Fig \ref{fig:3}. As discussed in Sec. \ref{Model}, these scales are directly related to the vacuum expectation value of the inflaton field $\Lambda \sim v_\phi$, which then influences the mass spectrum of both the scalar and Yukawa sectors. Next, we will explore the tree-level characteristics of the aforementioned $U(1)_{B-L}$ extension of the SM in light of the WIFI conjecture.

\subsection{The Scalar Sector}

The mass matrices of the scalar sector are computed from the bilinear terms in the potential given by Eq.~\eqref{Eq:Pot_1}, once the temperature of the thermal bath allows for the scalar fields to settle in their vacuum configuration (Eq.~\eqref{Eq:Min}). For the CP-even scalars such structure assumes the form
\begin{equation}
    M^2_{even} = \left( \begin{array}{ccc}
         2\lambda_h v^2_h & \lambda_1 v_h v_\sigma &  \lambda_2 v_h v_\phi\\
         \lambda_1 v_h v_\sigma & 2\lambda_\sigma v^2_\sigma & \lambda_3 v_\sigma v_\phi + \sqrt{2} m_{nh} v_\sigma \\
         \lambda_2 v_h v_\phi & \lambda_3 v_\sigma v_\phi + \sqrt{2} m_{nh} v_\sigma & 2\lambda_\phi v^2_\phi - \frac{m_{nh}v^2_\sigma}{\sqrt{2}v_\phi}
    \end{array} \right)\; ,
\end{equation}
obtained in the basis $(R_h,\, R_\sigma,\,R_\phi)$. 

The corresponding mass eigenstates are obtained following the usual methods of matrix diagonalization. It becomes useful, for this purpose, to attribute numerical values for the couplings and vevs of the scalar fields. In particular, the vev of the inflaton field is set by the cutoff scale for the DM production channel in Eq.~\eqref{Eq:CO}. As a representative example, let us consider the WI scenario of Figure \ref{fig:3} with a linear dissipation coefficient and a mass for the DM candidate particle of $1$ TeV. That implies $\Lambda = 4.5\times 10^{17}$ GeV, as follows from the observed DM abundance, and the vev $v_\phi \simeq 6.4 \times 10^{17}$ GeV. We will also set the vev of the $\sigma$ field to an intermediate energy scale, $v_\phi \sim 10^6$ GeV in order to address the WIFI hierarchy, but it is important to acknowledge that there is some freedom in the determination of this scale. As usual, the vev of the Higgs field follows from the electroweak breaking scale $v_h = 246$ GeV. In what concerns the couplings, the inflaton quartic coupling is fixed by the normalization of the scalar power spectrum to $\lambda_\phi \sim 3.3\times 10^{-15}$ (Sec. \ref{sec2}). Also, we impose $\lambda^2_2 \lesssim 4\lambda_h \lambda_\phi$ and $\lambda^2_3 \lesssim 4\lambda_\phi \lambda_\sigma$ in order to avoid a local minimum at the origin of field space. Here, we assume $\lambda_\sigma = \lambda_h = \lambda_1 \sim 10^{-1}$,  $\lambda_2 = \lambda_3 \sim 10^{-8}$.

For the benchmark points stated above, the diagonalization of the CP-even quadratic mass matrix results in three massive components, where the standard Higgs is the lightest one, $m_h \sim 10^2$ GeV, composed predominantly by $R_h$. One also obtains two heavy scalars, $m_\phi \simeq 5.2 \times 10^{10}$ GeV and $m_\sigma \simeq 4.3\times 10^{5}$ GeV, with the associated particles composed predominantly by $R_\phi$ and $R_\sigma$, respectively. 

Similarly, the quadratic mass matrix for the pseudo-scalars can be written in the basis $(I_h,\, I_\sigma,\,I_\phi)$, allowing for the structure
\begin{equation}
    M^2_{odd} = \left( \begin{array}{ccc}
         0 & 0 &  0\\
         0 & -2\sqrt{2}m_{nh} v_\phi & -\sqrt{2}m_{nh} v_\sigma \\
         0 & -\sqrt{2}m_{nh} v_\sigma & -\frac{m_{nh}}{\sqrt{2}}\frac{v^2_\sigma}{v_\phi}
    \end{array} \right),
\end{equation}
which reduces to a simple order two quadratic matrix. In this case, the eigenvalues are much more simple to compute, resulting in two massless particles, composed predominantly by $I_h$ and $I_\phi$. Those are the Nambu-Goldstone bosons absorbed as longitudinal components of the neutral gauge bosons $Z$ and $Z^\prime$. There is also a heavy pseudo-scalar, composed predominantly by $I_\sigma$, with quadratic mass
\begin{equation}
    m^{I\,2}_\sigma = -\frac{m_{nh}}{\sqrt{2}}\frac{4v^2_\phi + v^2_\sigma}{v_\phi}.
\end{equation}
Such mass scale is particularly susceptible to the value of the non-hermitian mass parameter, $m_{nh}$. Negative values are required in order to obtain a positive squared mass. For $10^{-10}\, \text{GeV} < |m_{nh}| < 10^{10} \, \text{GeV}$ one obtains $1.3\times 10^4 \, \text{GeV} \lesssim m^I_\sigma \lesssim 1.3\times 10^{14} \, \text{GeV}$ for the mass of this pseudo-scalar.

Finally, the quadratic mass terms for the charged scalars $(H^+,\, H^-)$ cancel completely, corresponding to the standard Nambu-Goldstone degrees of freedom described as longitudinal components of the charged weak bosons $W^\pm$.

\subsection{The Yukawa Sector}

The masses of the neutrino sector are obtained through the Yukawa terms in Eq.~\eqref{Eq:Yu_nu2}, which after the SSB process assume the matrix form,
\begin{eqnarray}
    M^\prime_\nu = \left(\begin{array}{ccc}
0 & Y^T_\nu \frac{v_h}{\sqrt{2}} & 0 \\ 
Y_\nu \frac{v_h}{\sqrt{2}} & Y_N \frac{v_\phi}{\sqrt{2}} & Y^T_M \frac{v_\sigma}{\sqrt{2}} \\
0 & Y_M \frac{v_\sigma}{\sqrt{2}} & \frac{v^4_\sigma}{M^3_\mu}\end{array}\right).
\end{eqnarray}
written in the basis $(\nu_L,\, N^c, \, \chi^c)$. One may assume, for simplicity, that the Yukawa matrices are diagonal and degenerated, $Y^{ij}_{\nu,N,M} = C_{\nu,N,M}\delta^{ij}$ and, for the sake of naturalness, that $C_{\nu,N,M} \gtrsim 10^{-3}$.  The vevs for the scalar fields are fixed, as before, to $v_h = 246$ GeV and $v_\sigma \simeq 10^6$ GeV, while $v_\phi$  is adjusted according to $\Lambda$ (see Eq. \eqref{Eq:CO}). Finally, the scale $M_\mu$ raises from loop diagrams where the sterile fermions are integrated out in the internal lines. In principle, it is possible to tune this parameter as to obtain the right hierarchy for the ISS mechanism and the desired scale of mass for the DM candidate. We considered $v^4_\sigma/M^3_\mu \sim 1$ GeV and $1$ TeV in our computations. 

The diagonalization of the $9\times 9$ mass matrix results in three standard neutrinos composed predominantly by $\nu_L$, three right-handed neutrinos composed predominantly by $N$, and three sterile fermions composed predominantly by $\chi$. The Yukawa couplings are fixed to $Y_\nu \sim 1$, $Y_M \sim 10^{-1}$ and $Y_N \sim 10^{-2}$ for the linear case, while $Y_\nu \sim 1$, $Y_M \sim 10^{-1}$ and $Y_N \sim 10^{-1}$ for the cubic. It is prudent to mention that, without the Majorana mass terms for right-handed neutrinos, $\bar{N}^c \mu_N N$, the 6 sterile fermions would acquire degenerate masses proportional to $M \sim v_\sigma$ and in turn would break the WIFI hierarchy $m_\chi < T < \Lambda$. The DM candidate mass is particularly susceptible to changes in $M_\mu$, while the standard and right-handed neutrino masses, $\nu$ and $N$, are dictated by a combination of the Yukawa factors $Y_\nu$, $Y_N$, and $Y_M$.

In Table \ref{tab:1}, we show the main quantities of our WIFI construction, for 1 GeV and 1 TeV DM particles, and both the linear and cubic dissipation coefficients. We display the average temperature $\bar T$ of the thermal bath sustained during WI, the cutoff scale $\Lambda$, the inflaton quartic coupling $\lambda$, the vevs of the inflaton and $\sigma$ fields, $v_\phi, v_\sigma$, and the masses of right-handed neutrinos and active neutrinos $m_N, m_\nu$. Although we have discussed practical examples, the results of this Section can be easily generalized to other values of DM masses. Changing the DM mass or the temperature dependence of the WI dissipation coefficient results in a different cutoff scale $\Lambda$, which, in turn, sets the vev of the inflaton field $v_\phi$. Therefore, one can obtain the masses for the active SM neutrinos in the sub-eV scale and heavy right-handed neutrinos quite naturally, by means of the high cutoff scale. The latter is required to be larger than the temperature of the WI bath, which is easily achievable, irrespective of the DM mass.

\begin{table*}
\centering
\begin{tabular}{|c|c|c|c|c|}
\hline
$m_\chi$ & \multicolumn{2}{|c|}{1 GeV} & \multicolumn{2}{|c|}{1 TeV}\\\hline\hline
& Linear & Cubic & Linear & Cubic \\\hline
$\bar T$ & $4.6\times 10^{14}$ GeV & $6.8\times 10^{14}$ GeV & $4.6\times 10^{14}$ GeV & $6.8\times 10^{14}$ GeV \\\hline
$\Lambda$ & $8.1 \times 10^{16}$ GeV & $7.6\times 10^{15}$ GeV & $4.5\times 10^{17}$ GeV & $4.2\times 10^{16}$ GeV \\\hline
$\lambda$ & $3.3 \times 10^{-15}$ & $2.7 \times 10^{-14}$ & $3.3 \times 10^{-15}$ & $2.7 \times 10^{-14}$ \\\hline
$v_\phi$ & $ 1.1 \times 10^{17}$ GeV & $ 1.1 \times 10^{16}$ GeV & $ 6.4 \times 10^{17}$ GeV & $ 5.9 \times 10^{16}$ GeV\\\hline
$v_\sigma$ & $ 10^{6}$ GeV & $ 10^{6}$ GeV & $ 10^{6}$ GeV & $ 10^{6}$ GeV \\\hline
$m_N$ & $8.1 \times 10^{14}$ GeV & $7.6 \times 10^{14}$ GeV & $4.5 \times 10^{15}$ GeV & $4.2 \times 10^{15}$ GeV\\\hline
$m_\nu$ & $3.7 \times 10^{-2}$ eV& $4.0 \times 10^{-2}$ eV & $6.7 \times 10^{-3}$ eV& $7.2\times 10^{-3}$ eV\\\hline

\end{tabular}
\caption{Numerical values for the main quantities of our WIFI model, for a 1 TeV and 1 GeV DM particle, as well as the linear and cubic WI dissipation coefficients.}
\label{tab:1}
\end{table*}

\subsection{The Dark Matter Candidate}

In order to further explore the viability of the model in accommodating a broader range of DM masses, we investigate the impact of the cutoff scale $\Lambda$ on the DM portal. The correct DM relic abundance is governed by the $Z^\prime$ portal to SM fermions, including all quarks and leptons (see Figure~\ref{sec3:diag}). Due to the effective interactions at a temperature below the mediator mass scale, DM is produced through the WIFI mechanism with $n=2$ (Figure~\ref{fig:3}).

The cutoff scale is related to the mass of the new gauge boson \cite{Elahi:2014fsa},
\begin{equation}
    \frac{1}{\Lambda^2} = \frac{Y^q_{\text{\tiny{B-L}}}Y^\chi_{\text{\tiny{B-L}}}g^2_{\text{\tiny{B-L}}}}{m^2_{Z^\prime}} \propto \frac{1}{v^{2}_\phi}, \label{Eq:CO}
\end{equation}
where $Y^q_{\text{\tiny{B-L}}}$ and $Y^\chi_{\text{\tiny{B-L}}}$ are the $B-L$ charges of the scattering particles in the thermal bath and of DM, respectively. Despite the reasonably low mass scale of the DM candidate, its interactions with the SM particles are suppressed by the energy scale of the $B-L$ symmetry breaking, $\Lambda \sim v_\phi$, which translates into difficulty in observing such a component directly ~\cite{Liu:2022zgu,LZ:2023lvz,XENON:2023cxc,XENON:2024wpa,LZ:2022lsv} or indirectly \cite{Fermi-LAT:2016uux,VERITAS:2017tif,CTAConsortium:2017dvg,CTA:2020qlo,Viana:2019ucn,MAGIC:2021mog,LHAASO:2022yxw,HESS:2022ygk,Calore:2022stf,HAWC:2023owv,IceCube:2023ies} by detection experiments.

\begin{figure*}[t]
\centering
\includegraphics[width=1\columnwidth]{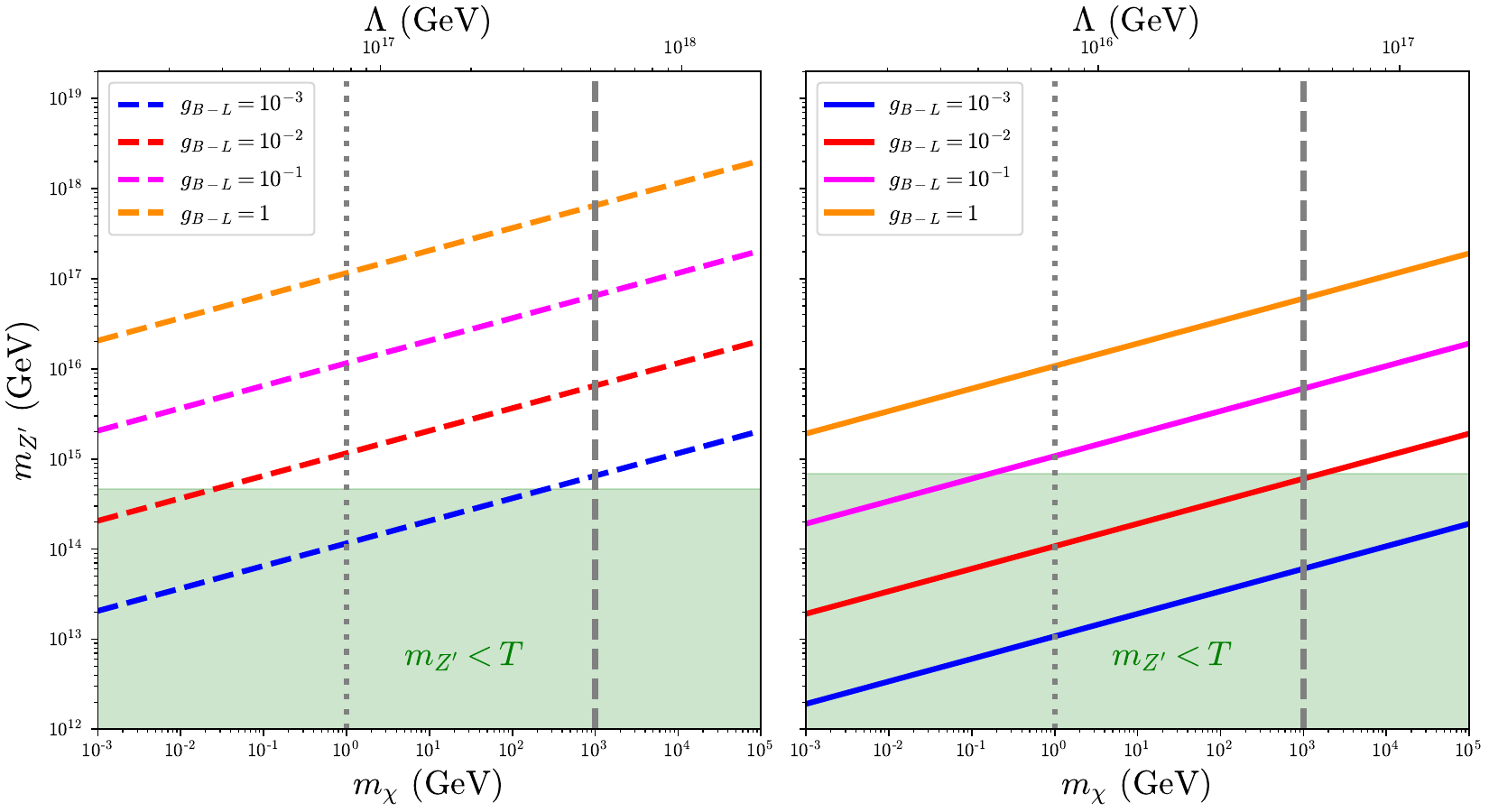}
\caption{The $Z^\prime$ mass as a function of the DM mass for representative values of the $B-L$ gauge coupling. On the left (right) we show the results for a linear (cubic) WI dissipation coefficient. The horizontal scale on the top of each graph displays the cutoff $\Lambda$ for each value of DM mass. The dotted and dashed gray vertical lines correspond to the 1 GeV and 1 TeV cases, respectively, discussed in Fig. \ref{fig:3}. The shaded green region, where $m_{Z^\prime} < T$, is excluded (see the text for details).}
\label{fig:4}
\end{figure*}

In Figure~\ref{fig:4}, we present the viable parameter space to address the UV freeze-in production of DM during warm inflation within our model. To this end, we choose four values for the $B-L$ gauge coupling ($g_{\text{\tiny{B-L}}}$), spanning over a wide range of DM ($m_\chi$) and the $B-L$ gauge boson ($m_{Z^\prime}$) masses. The main interactions that drive the production of DM in the very early universe are provided by the annihilation of SM fermions, produced through dissipation of the inflaton field during WI, into DM particles. The shaded green region is excluded by the constraint of a $Z^\prime$ mass higher than the WI bath temperature, in order to stay inside the validity of the effective DM-bath description and to avoid sizable thermal corrections to the inflaton potential (Sec. \ref{Corrections}). In particular, lowering the DM mass, which implies in a lower cutoff scale $\Lambda$, results in a lower value of the $Z^\prime$ boson mass, according to the relation in Eq.~\eqref{Eq:CO}. Therefore, for lower DM masses, one has to increase the $B-L$ gauge coupling to accomplish the WIFI mechanism, in order to evade the $m_{Z^\prime} < T$ region. This effect is more pronounced in WI with the cubic dissipation coefficient than with the linear one, since the former results in lower values of $\Lambda$. Also, the whole DM relic density is produced before the onset of the RD period (Figure \ref{fig:3}).

\section{Conclusions}\label{sec5}

In this work, we have constructed a realization of warm inflation freeze-in for DM production \cite{Freese:2024ogj}, in the context of the $U(1)_{B-L}$ gauge extension of the SM and the inverse seesaw mechanism for the generation of neutrino masses. In particular, we analyzed WI scenarios with a quartic potential in the weak dissipative regime ($\log Q_\star = -2$), in which the sustained thermal bath has a linear and a cubic dependence on the bath temperature. In our model, the DM-bath effective interaction is mediated by a non-renormalizable operator of mass dimension 6, in the framework of ultraviolet freeze-in \cite{Elahi:2014fsa}, which is equivalent of taking the exponent $n = 2$ in the Boltzmann equation for DM (Eq.~\eqref{boltzmann}). 

To demonstrate the DM production and the particle physics phenomenology of our WIFI construction of Sec. \ref{Model}, we considered the cases of 1 GeV and 1 TeV DM particles, which then assigns high scales to the cutoff $\Lambda$ of the effective DM-bath interaction. This high energy scale is dynamically generated through the spontaneous breaking of the $B-L$ gauge symmetry. The phase transitions are completed by the subsequent breaking of the discrete $(-1)^L$ and electroweak symmetries, allowing the model components to realize the ISS mechanism. The spectrum of the model comprehends six additional sterile fermions, where the lightest one is our DM candidate, a new gauge boson, two new heavy scalars and a pseudo-scalar. The latter acquires a mass scale proportional to the non-Hermitian free mass parameter, which translates into the possibility of a relatively light particle and possible limits may arise from missing energy in colliders \cite{ParticleDataGroup:2024cfk}. 

Our main results are summarized in Figure \ref{fig:3} and Table \ref{tab:1}. In the figure, we show the WIFI production of DM for both the linear and cubic dissipation coefficients and two representative examples of DM mass, 1 GeV and 1 TeV. The table provides the main parameters of our particle physics model of Sec. \ref{Model}, within the WIFI mechanism scenarios depicted in Figure \ref{fig:3}. As shown in Figure \ref{fig:4}, DM production is also possible for a wide range of DM masses and $B-L$ gauge couplings, although lighter DM particles require larger couplings in order to stay in the effective interaction regime. We have demonstrated that the WIFI mechanism is indeed feasible within minimal extensions of the SM, while also providing a hierarchy of scales that make it compatible with the seesaw mechanism for the generation of neutrino masses.

\acknowledgments
We thank Arjun Berera and Rudnei Ramos for the valuable discussions. RdS is supported by Coordena\c{c}\~ao de Aperfei\c{c}oamento de Pessoal de N\'ivel Superior (CAPES). JGR acknowledges financial support from Programa de Capacita\c{c}\~ao Institucional do Observat\'orio Nacional (PCI/ON/MCTI). CS is supported by Conselho Nacional de Desenvolvimento Cient\'ifico e Tecnológico (CNPq) grant No. 151364/2024-9 and by the S\~{a}o Paulo Research Foundation (FAPESP) through grant number 2021/01089-1. FBMS is supported by Conselho Nacional de Desenvolvimento Científico e Tecnológico (CNPq) grant No. 151554/2024-2. JA is supported by CNPq grant No. 307683/2022-2 and Funda\c{c}\~ao de Amparo \`a Pesquisa do Estado do Rio de Janeiro (FAPERJ) grant No. 259610 (2021). We also acknowledge the use of the High Performance Data Center
(CPDON) at the Observat\'orio Nacional for providing the computational facilities to run our
analyses.


\bibliographystyle{JHEP}
\bibliography{cas-refs}



\end{document}